# Structure, Dynamics and Themodynamics of a metal chiral surface: Cu(532)

Abdelkader Kara[*] and Talat S. Rahman

Department of Physics, Cardwell Hall, Kansas State University, Manhattan KS 66506

**Abstract**

The structure, vibrational dynamics and thermodynamics of a chiral surface, Cu(532), has been calculated using a local approach and the harmonic approximation, with interatomic potentials based on the embedded atom method. The relaxation of atomic positions to the optimum configuration results in a complex relaxation pattern with strong contractions in the bond length of atoms near the kink and the step site and an equivalently large expansion near the least under-coordinated surface atoms. The low coordination of the atoms on the surface affects substantially the vibrational dynamics and thermodynamics of this system. The local vibrational density of states show a deviation from the bulk behavior that persist down to the 10th layer resulting in a substantial contribution of the vibrational entropy to the excess free energy amounting to about 90 meV per unit cell at 300K.

**KEY WORDS:** chiral; stepped surfaces; vibrational entropy; thermodynamics

# Introduction

Since the middle of the 19th century, chirality was seen as one of the important molecular properties. Biological molecules maybe chiral with different enantiomers presenting opposite effects (one being used to cure and the other may cause harm). The fact that in nature, most of the chiral molecules exist only in one of the two possible forms or enantiomers, is by itself sufficient to excite scientific curiosity. Since the pharmaceutically fabricated chiral molecules exist in both forms, it is of great technological importance to design efficient enantio-selectors that will retain only the desired enantiomer. Since the two enantiomers of a chiral molecule have the same base atomic structure, their electronic structure is the same and they only differ in the sequence (clock/counter-clock) in which the atomic structure is formed. It is hence logical to use templates/substrates that are naturally or by design chiral [1-3].

For several decades, surface science community has been dealing with detailed analysis of heterogeneous catalysis and regions of low coordination, on the surface, have been often spotted as responsible for the onset of catalytic reactivity. Regularly kinked surfaces, which are inherently chiral, have been recently considered for enantio-selectivity [2,3]. Studies of chiral metal surfaces of Cu and Pt have also been the subject of recent studies [1,2,4,5]. It appears in *ab initio* calculations of Power and Sholl [4], that the difference in the energetics between relaxed and un-relaxed chiral surfaces of Pt may be important in the evaluation of the difference in adsorption energies of the two different enantiomers of a chiral molecule. These differences in energies are in the range of a few tenths of a kcal/mol or about a few tens of meV. For such small differences in the energetics the question of importance of thermal effects and of vibrational entropy becomes relevant in the considerations of the free energy.

In this paper, we present results for the structure, dynamics and thermodynamics of a chiral metal surface Cu(532) with the aim of extracting the contribution of vibrational entropy to considerations of surface energetics. In the next section, we present the atomic geometry and theoretical details. In section III we will discuss our results and conclusions.

## System Geometry and Theoretical Details

The system studied here is Cu(532), a kinked surface formed from a vicinal surface of Cu(111). By nature, all kinked surfaces are chiral in the sense that there is no mirror image of the surface that coincides with the surface. Since a kink is at the intersection of three planes, (here (111), (100) and (110)) the arrangement of these three planes around the corner classifies the surface as being of the R or the S type (here we follow the notations that has been discussed in details by Power and Sholl [4] and Ahmadi *et al* [6]). In Fig.1 we show a hard sphere model of $Cu(532)^S$ and $Cu(532)^R$. Note that the "S" surface corresponds to a counter-clockwise sequence of (111)-(100)-(110) while the "R" refers to a clockwise one. Note also that on this surface, the step separating two kinks is formed by two atoms forming a (100) micro-facet, with a 3-atom-wide terrace. The surface unit cell consists of eight atoms with coordination ranging between 6 and 11. The ninth atom is a neighbor of the kink atom with coordination twelve, labeled here BNN (bulk nearest neighbor). We shall see here as in our earlier work [8,12,13] that the BNN plays an important role in determining the characteristics of this system.

In Fig. 2, we show a top view of this surface along with the coordination of atoms on layers 1 (kink) to nine (BNN) (6, 7, 8, 9, 9, 10, 10, 11, and 12).

The interaction between Cu atoms is described by a many-body, embedded atom method (EAM) [7] potential. We have extensively used this potential to study the structure and dynamics of several low and high Miller index surfaces of Cu, Ag, Ni and Pd [8]. For Cu(532), after constructing the system in the bulk truncated positions, we apply standard conjugate gradient procedure to determine the minimum energy configuration. To calculate the vibrational density of states we use a real space Green's function method [9].

The advantage of this method is that one calculates the total density of states with no a-priori choice of wave vector, as would be the case in calculations based on the wave-vector or 'k-space'. Also, with some effort, one can get the polarization of the vibrational modes from the imaginary part of the columns of the Green's functions associated with the system at hand. This method exploits the fact that for a system with a finite range of interatomic interactions, the force constant matrix can always be written in a block tridiagonal form [9] in which the submatrices along the diagonal represent interactions between atoms within a chosen local region and the sub-matrices along the 'off-diagonal' correspond to interactions between neighboring localities. Thus an infinite/semi-infinite system is converted quite naturally into an infinite/semi-infinite set of local regions. There is hence no a-priori truncation in the system size, as would be the case for matrix diagonalization methods based on k-space. The real space Green's function method also has an advantage over the familiar 'continued fraction' method [10] as it does not involve truncation schemes to determine the recursion coefficients, rather a more general and simpler recursive scheme is applied [11]. The vibrational density of states corresponding to locality "l" is related to the trace of the Green's function by the well known relation

$$g_l(\omega^2) = -\frac{1}{3n_l\pi} \varepsilon \underset{0}{\lim} \, \mathrm{Im} Tr\left[G_{ll}(\omega^2+i\varepsilon)\right]$$

$$N_l(\omega) = 2\omega g_l(\omega^2)$$

where $G_{ll}$ is the Green's function sub-matrix associated with locality l and $n_l$ is the number of atoms in this locality. The Cu(532) unit cell contains 8 atoms (with coordination between 6 and 11) and the cut off radius of the EAM for Cu is about 5 Å, we choose the first locality to contain atoms in layers 1 to 40. This ensures a block-tri-diagonal for of the force constant matrix. Once the local vibrational density of states is calculated, we can determine easily the local thermodynamic properties of the system, in the harmonic approximation. The contribution of the local vibrational free energy $F_{loc}^{vib}$, the local vibrational entropy $S_{loc}^{vib}$, and the local mean square vibrational amplitude are thus given by:

$$F_{loc}^{vib} = k_B T \int_0^\infty \ln(2\sinh(x)) N_l(\omega) d\omega$$

$$S_{loc}^{vib} = k_B \int_0^\infty (x\coth(x) - \ln(2\sinh(x))) N_l(\omega) d\omega$$

$$\langle u_l^2 \rangle = (\hbar/2M) \int_0^\infty \frac{1}{\omega} \coth(x) N_l(\omega) d\omega$$

with x = (ℏω/2 $k_B$T), $k_B$ the Boltzman constant, T the temperature and M the atomic mass.

## Results and discussion

We start this section by presenting our results of the atomic relaxation of Cu(532). The general trend observed in the relaxation of vicinal (stepped) surfaces is applicable for this system as well. In particular, the lowest coordinated atoms (those at the kink and step sites) experience a large contraction while atoms that are the least under-coordinated experience a large expansion [8]. From Fig. 3, we note that the first four interlayer separations experience a contraction (as compared to the bulk one) of 10% or more while $\Delta d_{78}$ and $\Delta d_{89}$ show an expansion of about 13% followed by contractions of $\Delta d_{9,10}$ and $\Delta d_{10,11}$ of 9%. Even though the relaxation pattern show oscillations beyond $\Delta d_{10,11}$, these are relatively small and can be ignored.

As the loss of neighbors at the surface dictates a complex rearrangement of atomic positions, it also induces dramatic changes in the vibrational density of states of atoms at and near the surface and hence their thermodynamical properties, as we shall see. The calculated local density of states (LDOS) for the kink, step, BNN and bulk atoms of Cu(532) are shown in Fig. 4. We note that both the kink and the step atoms (with coordination 6 and 7, respectively) experience substantial enhancement of modes in the low frequency region accompanied with a significant reduction at the high frequency end. The large peak above the bulk band shown in the density of states of the BNN is related to the shortening of its bond length with the kink atom, as has already been found on several vicinal surfaces [12].

With the calculated vibrational density of states at hand, one can determine all contributions of vibrational entropy to the thermodynamics of the system. We focus the discussion here on the contribution of the under-coordinated atoms in the system, to the vibrational free energy, entropy and mean square vibrational amplitudes. As stated above, these quantities are calculated in the harmonic approximation which is expected to hold up to room temperature for Cu [14]. The plot in Fig. 5 of the local vibrational contribution of the kink, step, BNN and bulk atoms shows a much larger tendency of the step and kink atoms to lower the free energy for all temperatures. In Table I, we report the local vibrational contribution to the free energy ($F_{loc}^{vib}$) for atoms in layers one (kink) to nine (BNN). At room temperature, the contribution of the kink and step atoms is 18.73 meV and 18.20 meV, respectively, lower than that of the bulk atoms, which is a substantial difference. These values are very close to a step atom on Cu(211) (19.42 meV) reported earlier [13]. Vibrational contributions from all the atoms in the unit cell, to the free energy of Cu(532), sums up to an excess of 94.53 meV. This contribution is of the same order of magnitude as those involved in the differences in adsorption energies of enantiomers of a chiral molecule, and hence need be taken into account in considerations of the surface energetics.

The other thermodynamical quantity of interest that reflects the impact of the environment (coordination and relaxation) is the local vibrational entropy ($S_{loc}^{vib}$). In Table 2, we gather this value for atoms in layers 1 to 9, as well as, that for a bulk atom. As one would expect, the lower the coordination, the higher is the local vibrational entropy. As we have seen, low coordination results in a higher population of vibrational states at low frequencies yielding higher entropy. On the other hand, a slight "over-coordination" of the BNN atom, resulting from the inward relaxation around the kink atom, pushes its vibrational entropy lower than that of the bulk. We note also a slight non-monotonic decrease of the local vibrational entropy of atoms in layers 5 and 7, which reflects the local expansion of the interlayer separations. Finally, a quantity of direct relevance to the analysis of experimental structural data, the mean square vibrational amplitude associated with the top three layers atoms is plotted in Fig. 6 along with that associated with a bulk atom. We note from this figure that this quantity scales inversely with the coordination in agreement with the scaling of the local entropy. Note also that at room temperature, the mean square vibrational amplitude of the kink atom is an order of magnitude higher than that of the bulk. Since all thermodynamics quantities presented here were calculated in the harmonic approximation, the mean square vibrational amplitude scales linearly with the temperature at high temperatures, we report in Table 3 the slope of the mean square vibrational amplitude (taken at temperatures between 100K and 300K) for atoms in layers 1 to 9 along with that of the bulk. This slope decreases dramatically from 2.74 to 0.9 $Å^2/K$ $x10^{-4}$ as the atomic coordination varies from 6 (kink atom) to 12 (bulk atom) and should be useful for estimations of the Debye-Waller factors for the analysis of the decrease in the intensities of diffracted beams, as function of temperature, in the case of surface sensitive experimental techniques.

## Conclusions

The detailed study of the structure and vibrational dynamics and thermodynamics of the chiral surface Cu(532) shows that the relaxation pattern is governed by the bond-length/bond-order correlation. The equilibrium positions of the surface atoms tend to optimize their effective coordination number which results in an oscillatory relaxation pattern. The vibrational dynamics and thermodynamics of a kink atom are similar to those of a step atom showing saturation in the excess quantities with coordination. As a result of the high vibrational entropy associated with the low coordinated atoms on the chiral surface Cu(532), the contribution of the vibrational dynamics to the free energy is evaluated to be about 90 meV per surface unit cell (consisting of 8 atoms), in excess of the corresponding contribution from the bulk unit cell. This is a substantial contribution that should be taken into account when the energetics of the system are considered.

**Acknowledgement** This work was supported by the US-DOE under the grant DE-FG03-97ER45650.

**REFERNCES**

[*] Corresponding author: akara@ksu.edu

**Figure Caption:**

Figure 1: Geometry of an fcc(532) showing the two possible orientations.

Figure 2: Top view of an fcc(532) showing the coordination of the top 9 atoms.

Figure 3: Relaxation pattern for Cu(532).

Figure 4: Local vibrational density of states for the step, kink, BNN and bulk atoms.

Figure 5: Local contributions of the vibrational dynamics to the free energy for the kink, step, BNN and bulk atoms.

Figure 6: Local mean square vibrational amplitude for the atoms in the first 3 layers and the bulk.

Table 1 Local (layer) contribution of the vibrational free energy $F_{loc}^{vib}$ in meV.

| Layer # (coordination) | 0K | 100K | 300K |
|---|---|---|---|
| 1 (6) | 27.04 | 21.04 | -36.13 |
| 2 (7) | 26.69 | 21.01 | -35.60 |
| 3 (8) | 27.63 | 22.44 | -32.04 |
| 4 (9) | 28.70 | 23.90 | -29.66 |
| 5 (9) | 28.66 | 22.67 | -30.33 |
| 6 (10) | 30.23 | 25.76 | -26.57 |
| 7 (10) | 29.78 | 24.61 | -26.92 |
| 8 (11) | 31.00 | 28.56 | -19.45 |
| 9 (12) | 33.68 | 31.29 | -14.43 |
| Bulk (12) | 32.16 | 29.65 | -17.40 |

Table 2 Local (layer) vibrational entropy $S_{loc}^{vib}$ in $k_B$/atom.

| Layer # | 0K | 100K | 300K |
|---|---|---|---|
| 1 | 0.023 | 1.721 | 4.543 |
| 2 | 0.017 | 1.677 | 4.514 |
| 3 | 0.015 | 1.571 | 4.384 |
| 4 | 0.012 | 1.482 | 4.264 |
| 5 | 0.012 | 1.542 | 4.356 |
| 6 | 0.014 | 1.385 | 4.121 |
| 7 | 0.011 | 1.431 | 4.248 |
| 8 | 0.010 | 1.258 | 3.942 |
| 9 | 0.006 | 1.131 | 3.760 |
| bulk | 0.005 | 1.169 | 3.859 |

Table 3: Slope of the local mean square vibrational amplitude ($\text{Å}^2/\text{K} \times 10^{-4}$)

| Layer # | slope |
|---------|-------|
| 1 | 2.74 |
| 2 | 2.23 |
| 3 | 2.04 |
| 4 | 1.72 |
| 5 | 1.70 |
| 6 | 1.86 |
| 7 | 1.58 |
| 8 | 1.45 |
| 9 | 1.03 |
| bulk | 0.90 |

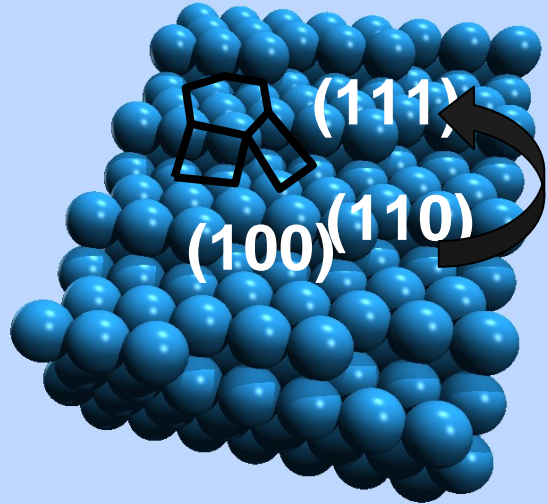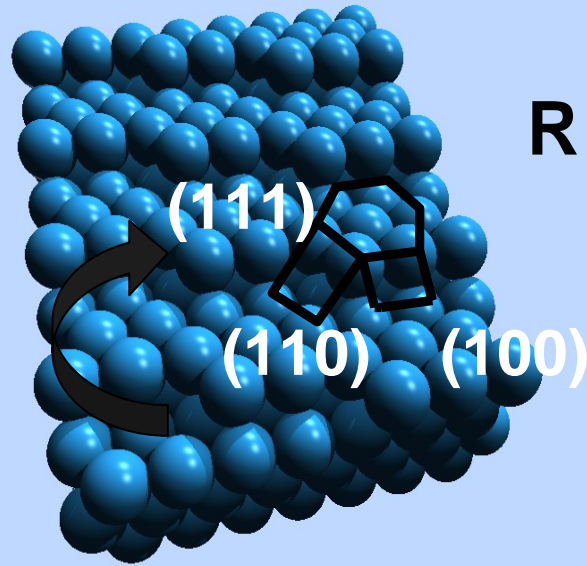

(kink) 6

(step) 7

(BNN)

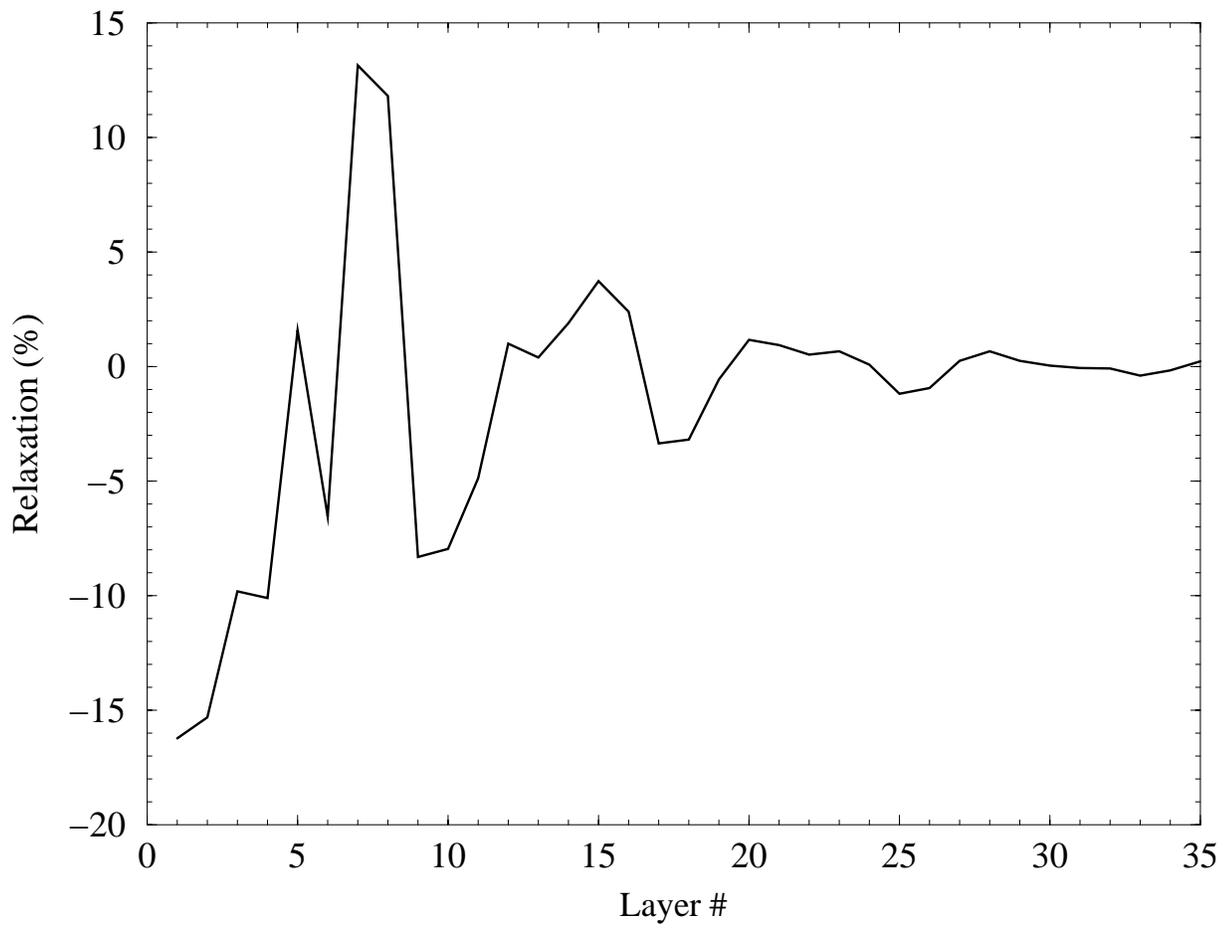

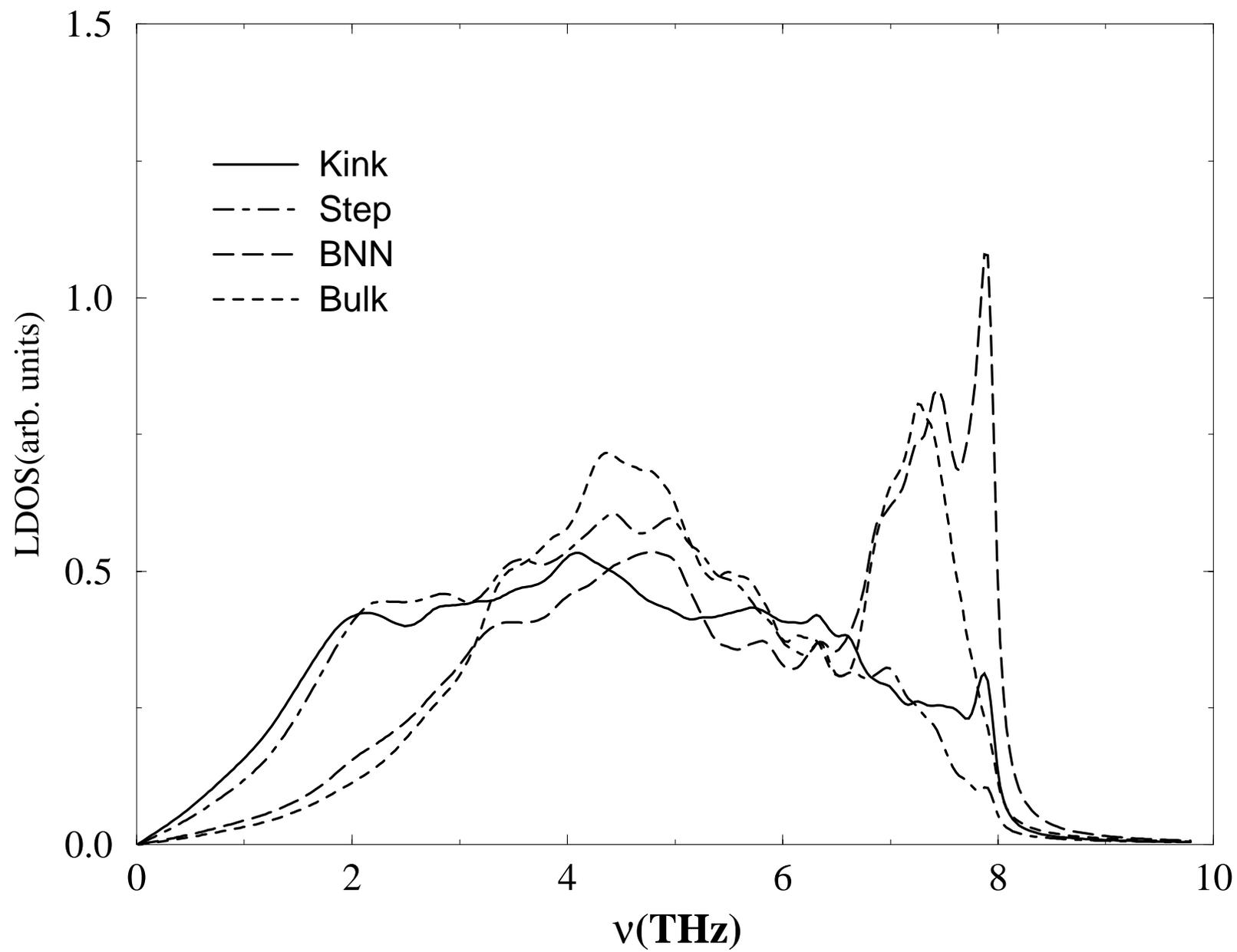

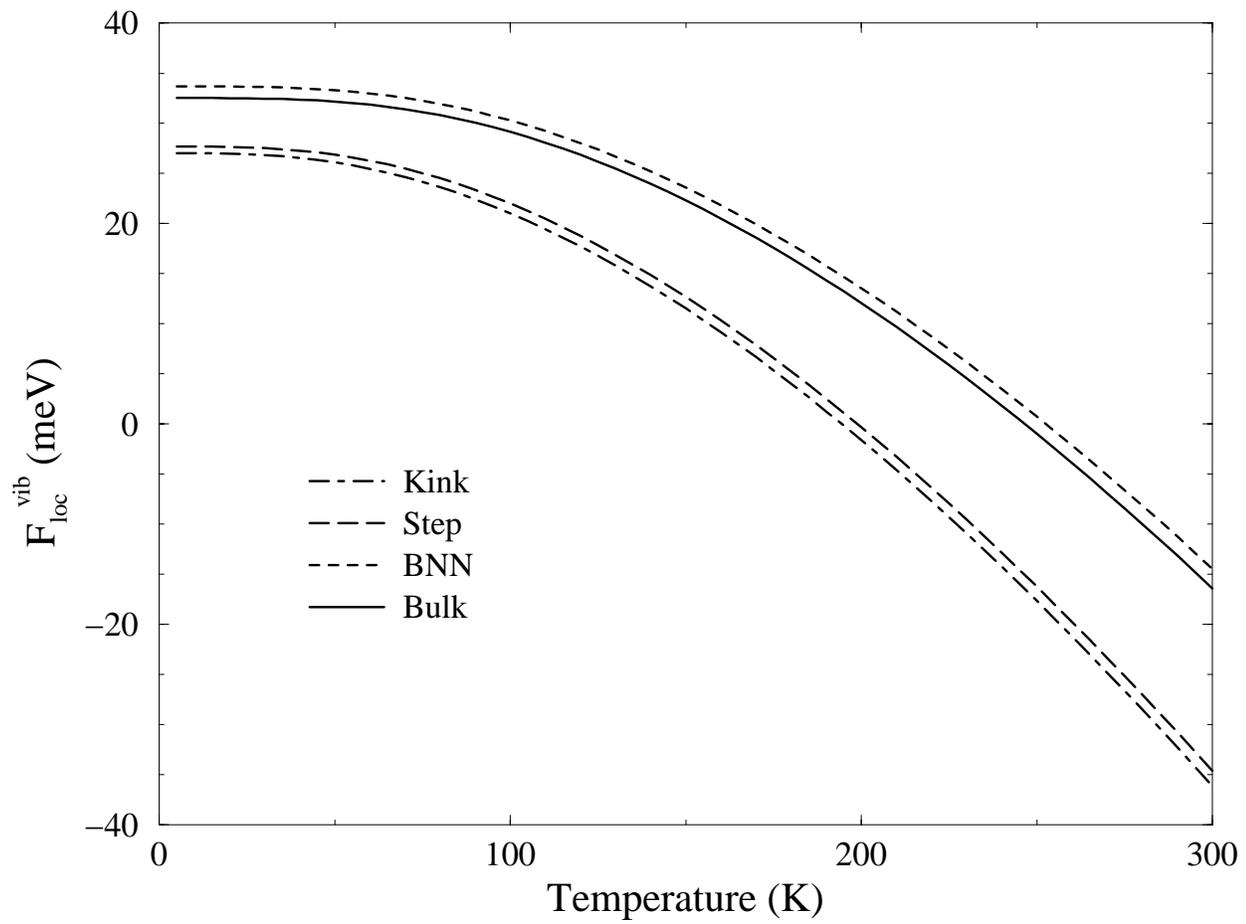

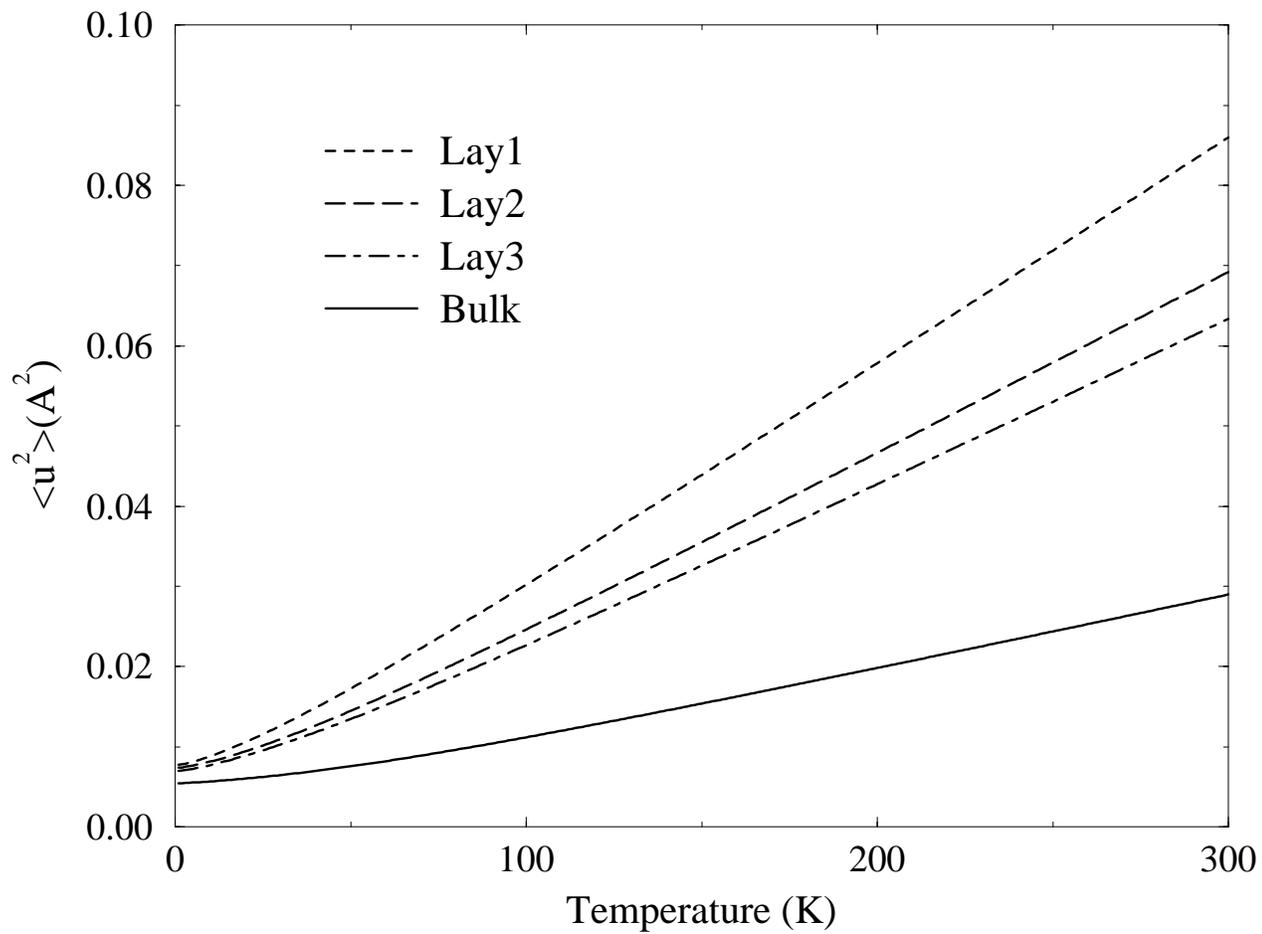